\documentclass[%
reprint,
superscriptaddress,
nofootinbib,
 amsmath,amssymb,
 aps,
 prl,
floatfix,
]{revtex4-2}

\usepackage{enumitem}
\usepackage{dsfont}
\usepackage{hyperref}
\usepackage{graphicx}
\usepackage{dcolumn}
\usepackage{bm}

\usepackage{epsfig}
\usepackage{amsmath}
\input{epsf}
\usepackage{psfrag}
\usepackage{xcolor}
\usepackage{pdfpages}
\usepackage[normalem]{ulem}
\usepackage{subcaption} 

\def\slash#1{\setbox0=\hbox{$#1$}  
   \dimen0=\wd0     
   \setbox1=\hbox{/} \dimen1=\wd1  
   \ifdim\dimen0>\dimen1   
      \rlap{\hbox to \dimen0{\hfil/\hfil}} 
      #1     
   \else     
      \rlap{\hbox to \dimen1{\hfil$#1$\hfil}} 
      /      
   \fi}      %

\makeatletter
\AtBeginDocument{\let\LS@rot\@undefined}
\makeatother
\newcommand{\bea}{\begin{eqnarray}}
\newcommand{\eea}{\end{eqnarray}}

\newcommand{\nn}{\nonumber}
\newcommand{\be}{\begin{equation}}
\newcommand{\ee}{\end{equation}}
\newcommand{\dd}{\mathrm{d}}
\newcommand{\e}{\mathrm{e}}
\newcommand{\bP}{\bm{P}}

\begin{document}

\preprint{XXX}

\title{NLO corrections and factorization for transverse single-spin asymmetries}

\author{Daniel Rein}
\email{da.rein@student.uni-tuebingen.de}
\author{Marc Schlegel}
 \email{marc.schlegel@uni-tuebingen.de}
\author{Patrick Tollk\"uhn}
 \email{patrick.tollkuehn@uni-tuebingen.de}
 \author{Werner Vogelsang}
 \email{werner.vogelsang@uni-tuebingen.de}
\affiliation{
Institute for Theoretical Physics, University of T\"ubingen, Auf der Morgenstelle 14, D-72076 T\"ubingen, Germany}

\date{\today}

\begin{abstract}
We present next-to-leading order QCD corrections for the cross sections for
$\ell p^\uparrow\to hX$, $\ell p^\uparrow\to {\mathrm{jet}}X$ with transversely polarized initial protons. 
These cross sections are known to be power-suppressed in QCD
and probe twist-3 parton correlation functions in the proton. Our calculation exhibits the full complexity of a
perturbative QCD analysis beyond leading power, involving in particular various derivatives of the 
parton correlation functions. We demonstrate that collinear factorization for these single-spin observables holds at one loop. 
We also present exploratory phenomenological results for the NLO single-spin asymmetry in $ep^\uparrow\to hX$ and 
compare to data from the HERMES experiment.

\end{abstract}

\maketitle

\textbf{\textit{Introduction.}} Single-transverse spin asymmetries (SSAs) are playing a pivotal role
for our understanding of nucleon structure and high-energy hadronic scattering in QCD. It was found that
they offer qualitatively new insights into the dynamics of quarks and gluons inside the proton, notably
thanks to their sensitivity to spin-orbit interactions, correlations among partons, and subtle phase
effects at the very heart of QCD as a gauge theory. At the same time -- and not surprisingly -- the theoretical analysis 
of SSAs has turned out to be remarkably complex. A centerpiece of research in this area has been 
the study of factorization theorems which, depending on the SSA observable, have to be formulated
either in terms of collinear momenta of partons in the proton, or additionally in terms of their transverse momenta. 
Thanks to remarkable theoretical achievements over the past few decades, there now is a mature
understanding of the basic frameworks of factorization for SSAs.  

In the present paper, we will address the SSA for the {\it single-inclusive process} $\ell p^\uparrow\to hX$,
with an incoming transversely polarized proton and a hadron $h$ produced at large transverse momentum $P_{h,T}$.
Note that we assume the scattered lepton to remain undetected. At sufficiently high energy, the hadron may also be 
replaced by an observed hadronic jet. The asymmetry is defined as
\be
A_{UT}\equiv \frac{E_h \frac{\dd \sigma}{\dd \bP_h }(\bm{S}_T)-E_h \frac{\dd \sigma}{\dd \bP_h }(-\bm{S}_T)}{E_h \frac{\dd \sigma}{\dd \bP_h }(\bm{S}_T)+E_h \frac{\dd \sigma}{\dd \bP_h }(-\bm{S}_T)}\,\equiv\,\frac{\Delta \sigma}{\overline\sigma} \,.\label{eq:SSA}
\ee
where $\bm{S}_T$ denotes the transverse proton spin vector and $(E_h,\bP_h)$ the hadron's four-momentum. 
The unexpectedly large values of $A_{UT}$ found in 
measurements for the single-inclusive reaction $pp^\uparrow\to hX$~\cite{Adams:1991rw,Krueger:1998hz,Allgower:2002qi,Adams:2003fx,Adler:2005in,Lee:2007zzh,Abelev:2008af,Arsene:2008mi,Adamczyk:2012xd,Adare:2013ekj,Adare:2014qzo,STAR:2020nnl} in fact marked the beginning of this field.~\footnote{
 The conceptual and kinematical similarity of the asymmetry \eqref{eq:SSA} to the corresponding one in $pp^\uparrow\to hX$ is the main motivation for us to study the single-inclusive process $\ell p^\uparrow\to hX$ instead of the more conventional semi-inclusive process $\ell p^\uparrow\to \ell hX$ (SIDIS).}

 A key property of single-inclusive reactions is that they are characterized by a single large momentum scale, here $P_{h,T}$. 
Therefore, collinear factorization is expected to provide the appropriate framework for the analysis. This should also apply
to the spin-dependent cross section in the numerator of $A_{UT}$, even though it is power-suppressed as 
$1/P_{h,T}$ relative to the unpolarized  one. Indeed, it was argued early 
on~\cite{Efremov:1981sh,Efremov:1983eb,Efremov:1984ip,Qiu:1991pp,Qiu:1991wg,Qiu:1998ia}  that $\Delta \sigma$ may be factorized
in terms of universal collinear ``twist-3'' hadronic matrix elements convoluted with partonic hard-scattering functions that are 
computable in QCD perturbation theory and are specific to the SSA under consideration. 
The hadronic matrix elements appear as correlation functions of partons in the proton, but also
in parton fragmentation to the observed hadron. 

An essential task for verifying factorization is to compute the next-to-leading order (NLO) corrections. In this paper we report
on the NLO calculation for the spin-dependent cross section for $\ell p^\uparrow\to hX$, focusing on contributions by initial-state
correlation functions. We select this process 
since it has the full 
complexity of twist-3 spin observables, but at the same time is much simpler than $pp^\uparrow\to hX$,
for which there are far more partonic channels. We note that there have been previous NLO calculations
for twist-3 SSAs, notably for Drell-Yan~\cite{Vogelsang:2009pj,Chen:2016dnp}, semi-inclusive 
DIS~\cite{Kang:2012ns,Dai:2014ala,Yoshida:2016tfh,Chen:2017lvx,Benic:2019zvg}, 
and hyperon production in $e^+e^-$ annihilation~\cite{Gamberg:2018fwy}. Although these calculations are most valuable, 
the analysis of the involved observables is considerably simpler compared to that for $\ell p^\uparrow\to hX$. 
This is manifested, in particular, by the absence of derivative contributions of the twist-3 correlation functions,
which for $\ell p^\uparrow\to hX$ are present already at lowest order because of the more complex momentum flow
for this process. This may seem a technicality, but is in fact at the core of the twist-3 formalism~\cite{Qiu:1991pp,Qiu:1991wg,Qiu:1998ia}.
We will demonstrate factorization for $\ell p^\uparrow\to hX$ at NLO, showing that all 
collinear divergences may be consistently absorbed into the twist-3 correlation functions, even in the presence
of derivative terms. We will also derive all finite pieces. At the same time, modeling the presently unknown correlation functions, 
we will also demonstrate that it is possible to carry out phenomenology for the NLO corrections.
All in all, we view our calculation as a breakthrough on the way to future full NLO calculations for  $pp^\uparrow\to hX$.

\textbf{\textit{Twist-3 correlation functions.}} 
In factorization of the single-spin cross section at twist-3, a key quantity is the correlation function
\bea
\Phi_{ij}^{\rho}(x,x^\prime)&=& \int_{-\infty}^\infty \tfrac{\dd\lambda}{2\pi} \int_{-\infty}^\infty \tfrac{\dd \mu}{2\pi} \,\e^{i\lambda x^\prime}\e^{i\mu (x-x^\prime)}\\
&&\hspace*{-1.7cm}\times\langle P,S|\,\bar{q}_j(0)\,[0,\mu n]\,ig\,G^{n\rho}(\mu n)\,[\mu n,\lambda n]\,q_i(\lambda n)\,|P,S\rangle .\label{eq:DefPhiF}\nn
\eea
It consists of ``dynamical'' matrix elements of two quark field components $q_i$, $q_j$ and a gluonic field strength tensor $G$, separated by Wilson lines denoted by $[\ldots]$. 
As indicated, the correlator depends on two light-cone momentum fractions $x$ and $x^\prime$. As a full matrix in Dirac space, it can be parameterized as~\cite{Kanazawa:2015ajw}
\begin{eqnarray}
\Phi^{\rho} =
\frac{M}{2}\left(i\epsilon^{Pn\rho S}\slash{P}F_{(qg)}^{\mathbf{1}}(x,x^\prime)- S_T^\rho\slash{P}\gamma_5F_{(qg)}^{\mathbf{5}}(x,x^\prime)\right),\;\;\;\;
\label{eq:DefPhiF1}
\end{eqnarray}
up to terms that are not relevant to this analysis. Here $M$ is the nucleon mass. Similar expressions may be written for three-gluon
matrix elements. 

\begin{figure}[h]
\vspace*{-4mm}
  \epsfig{figure=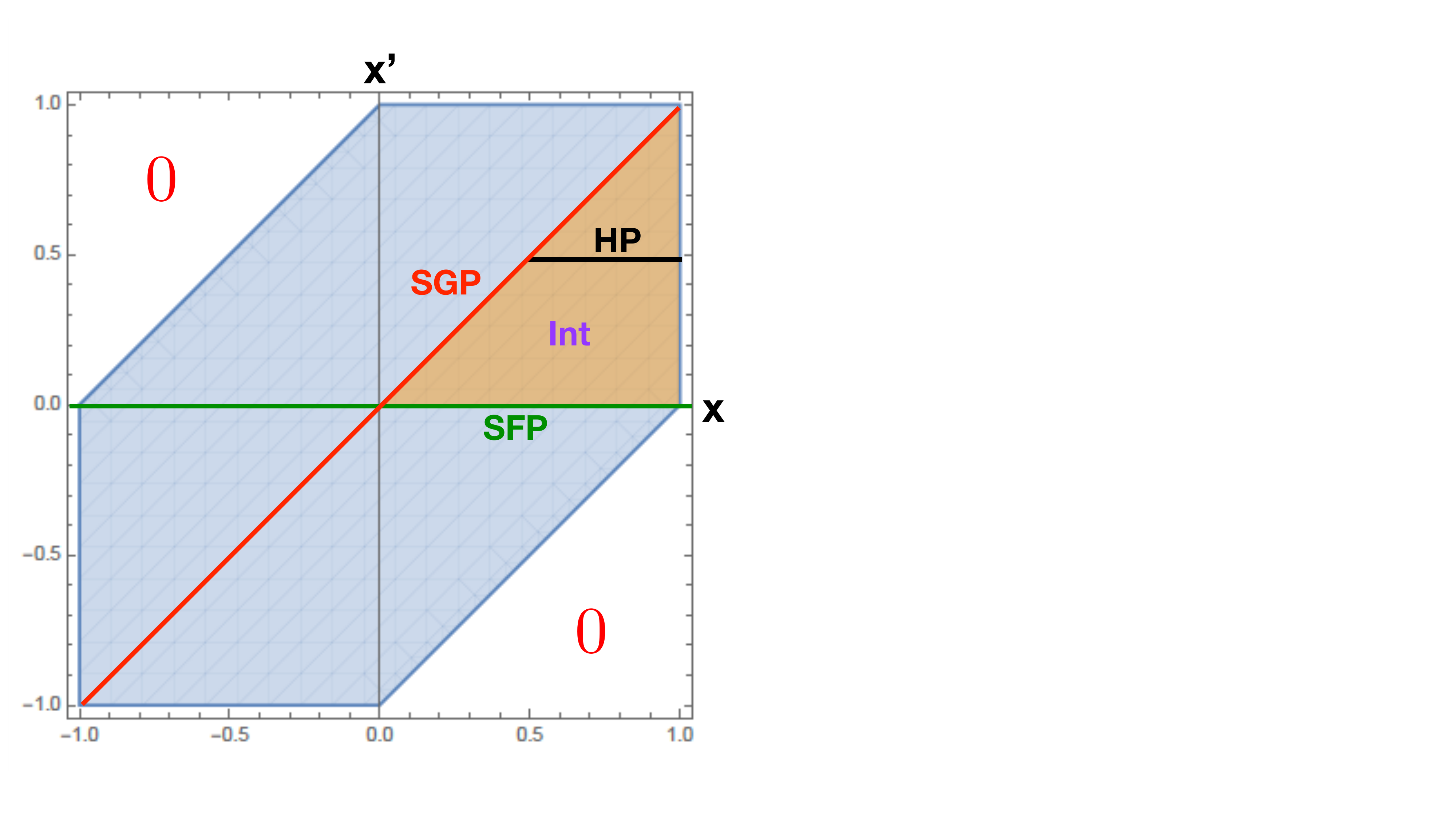,width=0.9\textwidth}
  \hspace*{2cm}
  \vspace*{-1cm}
  \caption{Region of support in momentum fractions $x,x'$ of $F_{(qg)}^{\mathbf{1,5}}(x,x^\prime)$.  }\label{fig1}
\end{figure}

The support of the functions $F_{(qg)}^{\mathbf{1,5}}(x,x^\prime)$ is given by the conditions $-1\le x,x^\prime \le 1$ and $|x-x^\prime|\le 1$. 
Figure~\ref{fig1} shows the ensuing region in the $x,x'$ plane. In general, the functions will enter a factorized cross section with an
integration over all kinematically accessible  $x$ and $x'$. However, as is well known, a non-vanishing single-spin asymmetry requires the presence of
a phase in the scattering process. Such a phase may be generated when a propagator (momentum $q$) in the partonic hard-scattering goes on-shell, via the relation 
\be\label{prop}
\mathfrak{Im}\frac{1}{q^2+i\varepsilon}=-i\pi\delta(q^2)\,.
\ee
Quite in general, the momentum $q$ will depend on the incoming partons' momentum fractions $x,x'$, leading to a restriction of the
region of $x,x'$ in which the functions $F_{(qg)}^{\mathbf{1,5}}(x,x^\prime)$ are probed. The following four cases occur in 
our calculation of the NLO hard-scattering function for the spin-dependent cross section:
\begin{itemize}
\item $x=x'$. For these contributions the gluon in the correlation function does not carry any longitudinal momentum. 
They are thus known as {\it soft-gluon pole} (SGP) contributions. The functions $F_{(qg)}^{\mathbf{1}}(x,x^\prime)$
appearing here are {\it Qiu-Sterman matrix elements} \cite{Qiu:1991pp}. 
\item $x^\prime=0$. Here one of the initial quarks has vanishing momentum fraction; these are {\it soft-fermion pole} (SFP) contributions.
\item $x$ is integrated and $x'$ is fixed at some kinematical point with $x'\neq x,0$. These are {\it hard-pole} (HP) contributions.
\item There are integrations over both $x$ and $x'$. Such contributions arise when $q$ in Eq.~(\ref{prop}) contains 
the momentum of an unobserved final-state parton which has to be integrated over via a phase space integral.
 We refer to them as {\it integral} (Int) contributions~\footnote{We note that for these contributions typically direct application of~(\ref{prop}) results in $q$ appearing in the arguments of the functions $F_{(qg)}^{\mathbf{1,5}}(x,x^\prime)$, which clearly is not desirable. It is easier in such cases to first carry out the integration over the momentum of the
 unobserved particle and take the imaginary part afterwards.}.
\end{itemize}

\twocolumngrid

\textbf{\textit{Factorized spin-dependent cross section.}} 
Using the hadronic Mandelstam variables $s=(l+P)^2$, $t=(P-P_h)^2$ and $u=(l-P_h)^2$ we introduce 
\be
x_0 \equiv \frac{1-v}{v} \frac{u}{t}, \quad  v_0 \equiv \frac{u}{t+u} ,\quad v_1 \equiv \frac{s+t}{s}\,.\label{eq:boundaries}
\ee
To NLO, the factorized spin-dependent cross section may be cast into the form (disregarding twist-3 fragmentation contributions)
\bea
E_h\frac{\dd \Delta \sigma}{\dd^{3} \mathbf{P}_h} & = & \sigma_0
\,\int_{v_0}^{v_1}\dd v\int_{x_0}^1\frac{\dd x}{x}\int_0^1\dd \zeta\label{eq:fact}\nn\\
&&\hspace*{-1.6cm}\times\sum_{\mathcal{P}} \sum_{(ij),k} \sum_{\mathbf{I}=\mathbf{1,5}} 
\hat{\sigma}_{(ij)\to k}^{\mathcal{P},\mathbf{I}}(v,w,\zeta)\, {\cal F}_{(ij)}^{\mathcal{P},\mathbf{I}}(x,\zeta x)\,D_k^h(z),\;\;\;
\eea
where $w=x_0/x$, $z=(1-v_1)/(1-v)$ and
\be
\sigma_0 \equiv \left(\frac{2 \alpha_{\mathrm{em}}^2}{-su}\right)\,\left(\frac{4\pi M\epsilon^{lPP_hS}}{-su}\right)\,,\label{eq:prefactor}
\ee
(with the fine structure constant $\alpha_{\mathrm{em}}$)
contains the dependence on the transverse polarization vector $\bm{S}_T$. The explicit appearance of the nucleon mass 
shows the typical power suppression of the single-spin cross section. The first sum in Eq.~(\ref{eq:fact}) runs over
the pole-type contributions ${\mathcal{P}}={\mathrm{SGP,HP,Int,SFP}}$ discussed above, while the second sum collects
all partonic channels with the partons in the pair $(ij)$ both entering from the proton, and parton $k$ fragmenting. 
The final sum takes into account non-chiral ($F_{(qg)}^{\mathbf{1}}$) and chiral ($F_{(qg)}^{\mathbf{5}}$) contributions. The $\hat{\sigma}_{(ij)\to k}^{\mathcal{P},\mathbf{I}}$ 
are the associated partonic hard-scattering cross-sections. They have the perturbative expansions
\be
\hat{\sigma}_{(ij)\to k}^{\mathcal{P},\mathbf{I}}\,=\,\hat{\sigma}_{(ij)\to k}^{\mathcal{P},\mathbf{I},{(0)}}
+\frac{\alpha_s}{2\pi}\,\hat{\sigma}_{(ij)\to k}^{\mathcal{P},\mathbf{I},(1)}\,+\,{\cal O}(\alpha_s^2)\,.
\ee
We report here on the calculation and results for the NLO terms $\hat{\sigma}_{(ij)\to k}^{\mathcal{P},\mathbf{I},(1)}$. 
The functions ${\cal F}_{(ij)}^{\mathcal{P},\mathbf{I}}$ in Eq.~(\ref{eq:fact})  
are derived from the twist-3 correlation functions $F_{(qg)}^{\mathbf{1,5}}(x,x^\prime)$ introduced in~(\ref{eq:DefPhiF}),
in a way that depends on the perturbative order of the calculation. The integration over $\zeta=x'/x$ in~(\ref{eq:fact}) is
non-trivial only for the integral contributions; all other contributions carry an explicit $\delta$-distribution that fixes $\zeta$. 
Finally, $D_k^h$ denotes a standard leading-twist fragmentation function for parton $k$ producing the observed hadron.
For $\ell p\to {\mathrm{jet}}X$ it is to be replaced by a suitable jet function, along with required modifications
of the hard-scattering functions. We note that the various functions in~(\ref{eq:fact}) are tied together by their
dependence on a factorization/renormalization scale $\mu$, which has been omitted for brevity. 

\textbf{\textit{Lowest-order contribution.}} 
At LO, only the channel $(qg)\to q$ contributes. Only an SGP contribution is present, and one readily 
finds~\cite{Kang:2011jw,Gamberg:2014eia,Kanazawa:2015ajw}
\bea
\hat{\sigma}_{(qg)\to q}^{\mathrm{SGP},\mathbf{1},{(0)}}(v,w,\zeta) &=&e_q^2\, \frac{1+v^2}{(1-v)^4}\,\delta(1-w)\,\delta(1-\zeta)\,.\nn\\
{\cal F}_{(qg)}^{\mathrm{SGP},\mathbf{1}}(x,x,\mu)\big|_{\mathrm{LO}}&=&\left(F_{(qg)}^{\mathbf{1}} - x\,\tfrac{\dd}{\dd x}F_{(qg)}^{\mathbf{1}}\right)(x,x,\mu).\;\;
\label{eq:LOpartonic}
\eea
As characteristic of an SGP contribution, there is a factor $\delta(1-\zeta)$, and the twist-3 correlation function is probed on the 
diagonal $x'=x$, implying also that only $F^{\mathbf{1}}_{(qg)}$ contributes. We also note the distribution $\delta(1-w)$, which 
expresses the fact that the recoiling final state consists of a single massless parton. One should keep in mind that, while the LO partonic cross
section only receives SGP contributions, the evolution of the twist-3 matrix elements $F^{\mathbf{1}}_{(qg)}(x,x,\mu)$
is sensitive also to HP and SFP configurations~\cite{Braun:2009mi}.

\begin{figure}[h]
  \epsfig{figure=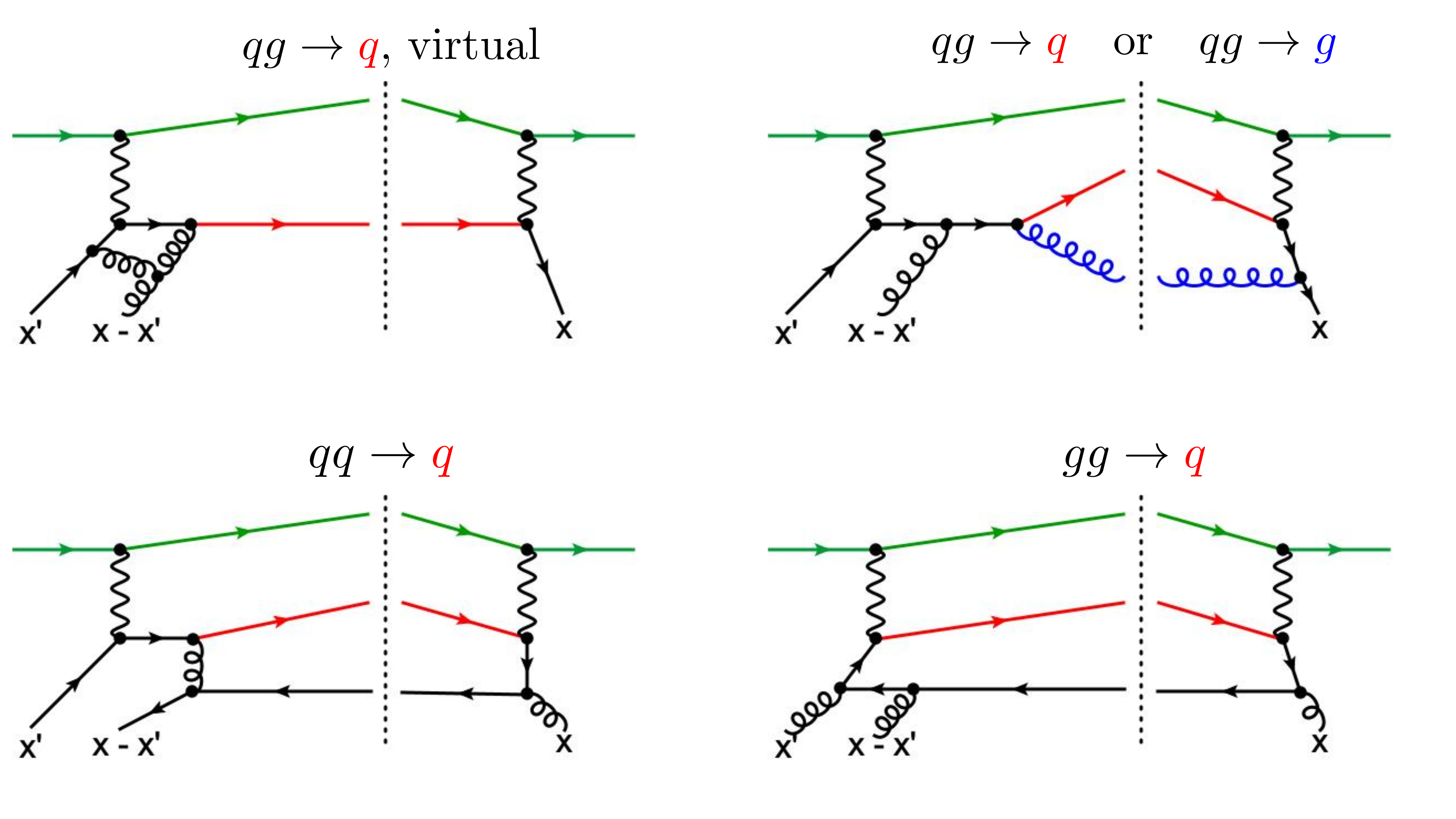,width=0.5\textwidth}
  \vspace*{-0.5cm}
  \caption{Sample NLO diagrams for the various partonic channels. For the $(qg)\to q$ channel we show virtual and real corrections separately. The $(qg)\to q$ and $(qg)\to g$ channels are generated by the same diagrams, but with different fragmenting partons (\textcolor{red}{quark} or \textcolor{blue}{gluon}).}\label{fig2}
\end{figure}

\textbf{\textit{Next-to-leading order calculation.}} Beyond LO, there are new partonic channels on top of $(qg)\to q$. 
These are generically denoted by $(qg)\to g$, $(qq)\to q$, 
and $(gg)\to q$
, where $q$ can variously also represent an antiquark and where quark
flavors in the initial and final states may differ. Sample NLO diagrams for the four channels are
shown in Fig.~\ref{fig2}. We work in $d=4-2\varepsilon$ dimensions to regularize soft and collinear divergencies.
We note that we set the lepton mass $m_l$ to zero. This simplifies the calculation but 
is known to generate spurious collinear singularities that are not present for the physical lepton mass. The proper way of 
subtracting these singularities and re-introducing the correct leading dependence on $m_l$ is well known and adopted here (see, for example,~\cite{Hinderer:2015hra}).

\twocolumngrid

The virtual corrections for $(qg)\to q$ exhibit both SGP and HP contributions, the latter arising from imaginary
parts of the loop integrals. The SGP contributions are proportional to the LO cross section and contain
a $1/\varepsilon^2$ divergence that cancels against an opposite term in the real-emission diagrams~\cite{Hinderer:2017bya}.

The cancelation of single $1/\varepsilon$ terms in the NLO calculation is much more subtle and complex. 
It involves judicious combination of all types of pole contributions arising in virtual and real diagrams, 
SGP, SFP, HP, and especially also {\it integral} contributions which themselves turn into SGP, SFP, HP terms 
on certain lines of their integration domain (see Fig.~\ref{fig1}). Furthermore, one even needs to exploit
the fact that the $(qq)\to q$ channel (see Fig.~\ref{fig2}(b)) probes the same correlation function
as that for $(qg)\to q$, just at different momentum fractions. This gives rise to a further SFP contribution 
that is needed to cancel $1/\varepsilon$ divergences in the SFP part of $(qg)\to q$. 

The final ingredient for arriving at a finite answer is the proper ($\overline{\mathrm{MS}}$) subtraction of collinear
divergencies corresponding to renormalization of the LO combination of twist-3 correlation functions
(see Eq.~(\ref{eq:LOpartonic})), $(F_{(qg)}^{\mathbf{1}} - x\,\dd F_{(qg)}^{\mathbf{1}}/{\dd x})(x,x,\mu)$.
The relevant subtraction terms may be obtained from the computations of evolution kernels of $F_{(qg)}^{\mathbf{1}}$~\cite{Braun:2009mi}. 
We note that the kernels themselves contain SGP, SFP, and HP contributions. 

The central result of this paper is that upon combination of all terms we arrive at finite results for all partonic 
channels contributing at NLO to the spin-dependent cross section~\cite{supplement}.
 This 
validates factorization at collinear twist-3 to order $\alpha_s$.
For single-inclusive jet production, our result presents the complete NLO answer; we
recall that for hadron production there will also be twist-3 fragmentation contributions that we do not
address here \cite{Gamberg:2014eia,Kanazawa:2015ajw}. 

The final expressions for the NLO corrections are lengthy and will be reported in a companion paper~\cite{Rein:2025qhe}.
Here we only present the leading double-logarithmic threshold contribution for $w\to 1$, which is entirely SGP and reads:
\be
\hat{\sigma}_{(qg)\to q}^{\mathrm{SGP},\mathbf{1},{(1)}}
\approx 8e_q^2\, C_F\,\frac{1+v^2}{(1-v)^4}\,\left(\frac{\ln(1-w)}{1-w}\right)_+\delta(1-\zeta).\label{eq:NLOpartonic}
\ee
with the usual ``plus'' prescription. It enters with the combination 
$(F_{(qg)}^{\mathbf{1}} - x\,\dd F_{(qg)}^{\mathbf{1}}/{\dd x})(x,x,\mu)$ and thus leads to a 
multiplicative correction to the LO cross section, with exactly the same coefficient as in the unpolarized
case~\cite{Hinderer:2015hra,Hinderer:2017bya}. This is expected because of its association with soft-gluon emission. We note that the
single-logarithmic correction $\propto 1/(1-w)_+$ differs from that in the unpolarized case, even in terms
of color factors. 

As mentioned above, our calculation also includes the $(gg)\to q$ channel, which is driven by
three-gluon twist-3 correlation functions. For the collinear subtraction in this channel we recover the three-gluon contributions to the evolution
of $F_{(qg)}^{\mathbf{1}}$ previously derived in~\cite{Kang:2008ey}. 

\begin{figure*}[h]
  \epsfig{figure=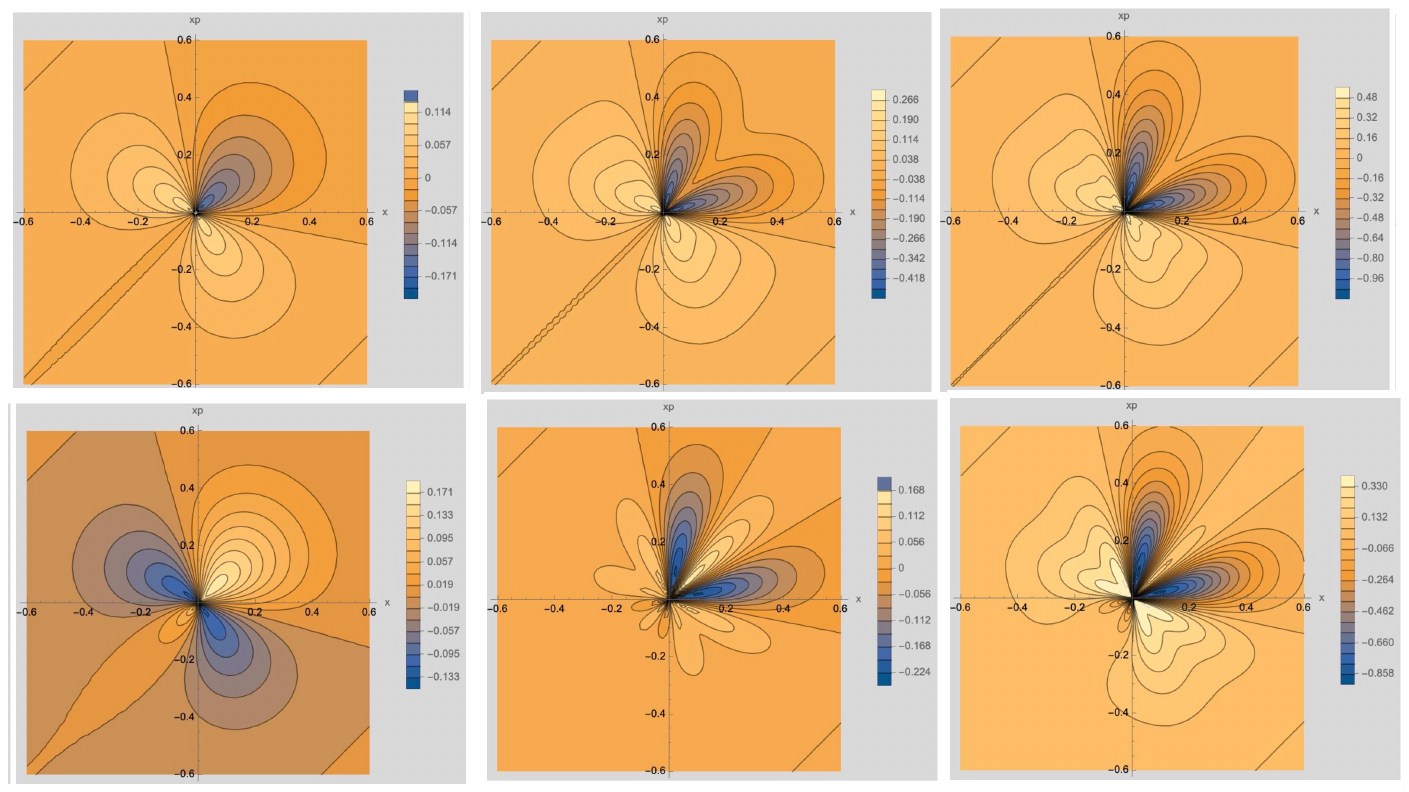,width=0.8\textwidth}
  \vspace*{-2.5cm}
  \caption{Contour plots of our model twist-3 $u$-quark function $F_{(ug)}^{\mathbf{1}}(x,x^\prime)$ (upper row) and $d$-quark function $F_{(dg)}^{\mathbf{1}}(x,x^\prime)$ (lower row) in the $x,x'$ plane.}\label{fig3}
\end{figure*}
\twocolumngrid

\textbf{\textit{Phenomenology of single-spin asymmetry at NLO.}}
We now present a few exploratory numerical results for the single-spin asymmetry in $ep^\uparrow \to\pi^\pm X$, 
adopting kinematics relevant for comparison to HERMES data \cite{Airapetian:2013bim} taken at c.m.s. energy $\sqrt{s}=7.25$~GeV. 
For the spin-dependent cross section we need input distributions for the quark-gluon-quark correlation functions $F_{(qg)}^{\mathbf{1,5}}$ for all flavors $q$, 
as well as for the three-gluon correlation functions. Since none of these functions are currently known, a realistic prediction is
not feasible. The only constraints are provided by the connection of the Qiu-Sterman functions to the first $k_\perp$ moments of the Sivers functions~\cite{Boer:2003cm}, 
$f_{1T}^{\perp (1),q}(x) = \pi F_{(qg)}^{\mathbf{1}}(x,x)$, along with $F_{(qg)}^{\mathbf{5}}(x,x)=0$, and lattice-QCD 
computations~\cite{Gockeler:2005vw,Burger:2021knd,Bickerton:2020hjo,Crawford:2024wzx} 
of the quantity $\int_{-1}^1\dd x\int_{x-1}^1\dd x^\prime F_{(qg)}^{\mathbf{1}}(x,x')$. For our numerical studies
we resort to a set of three simple models for $F_{(qg)}^{\mathbf{1,5}}(x,x')$ for up and down quarks, assuming all other multiparton correlation 
functions to vanish. Technically, we write the momentum fractions in polar coordinates, $x=r\cos(\varphi+\tfrac{\pi}{4})$, 
$x^\prime=r\sin(\varphi+\tfrac{\pi}{4})$, and then expand the $\varphi$ dependence of $F_{(qg)}^{\mathbf{1,5}}$ in terms
of a Fourier series, selecting three sets of values for the low Fourier modes.
Figure~\ref{fig3} shows the three resulting $u$- and $d$-quark functions $F_{(qg)}^{\mathbf{1}}$ as contour plots
in the $x,x'$ plane. Further details as well as a related plot of $F_{(qg)}^{\mathbf{5}}$ are given in our companion paper~\cite{Rein:2025qhe}. 
We stress that each one of the models in Fig.~\ref{fig3} respects the Sivers function and lattice constraints described above.
To compute the $k_\perp$ moments of the Sivers functions we adopt the analysis of Ref.~\cite{Anselmino:2008sga}. Note that for this exploratory study we do not take into account the proper scale evolution of the twist-3 correlation functions but only vary the scale of the unpolarized $f_1^q$ parton distributions underlying our model.
We then generate theoretical uncertainty bands by varying the scale in the range $1~{\mathrm{GeV}}\leq\mu\leq 2.2~{\mathrm{GeV}}$
corresponding to the boundaries of the HERMES $P_{\pi,T}$ bin,
using $\mu=\langle P_{\pi,T} \rangle\simeq 1.15$~GeV as the central value.
%
%
We use the pion fragmentation functions of Ref.~\cite{deFlorian:2014xna}. We recall
that we do not take into account any twist-3 fragmentation contributions. 
The unpolarized cross section in the denominator of the asymmetry is computed at NLO, using the results of~\cite{Hinderer:2015hra}
and the MSTW2008 parameterization~\cite{Martin:2009iq} for the unpolarized proton's parton distributions.

\begin{figure*}[h]

\hspace*{0cm}
  \epsfig{figure=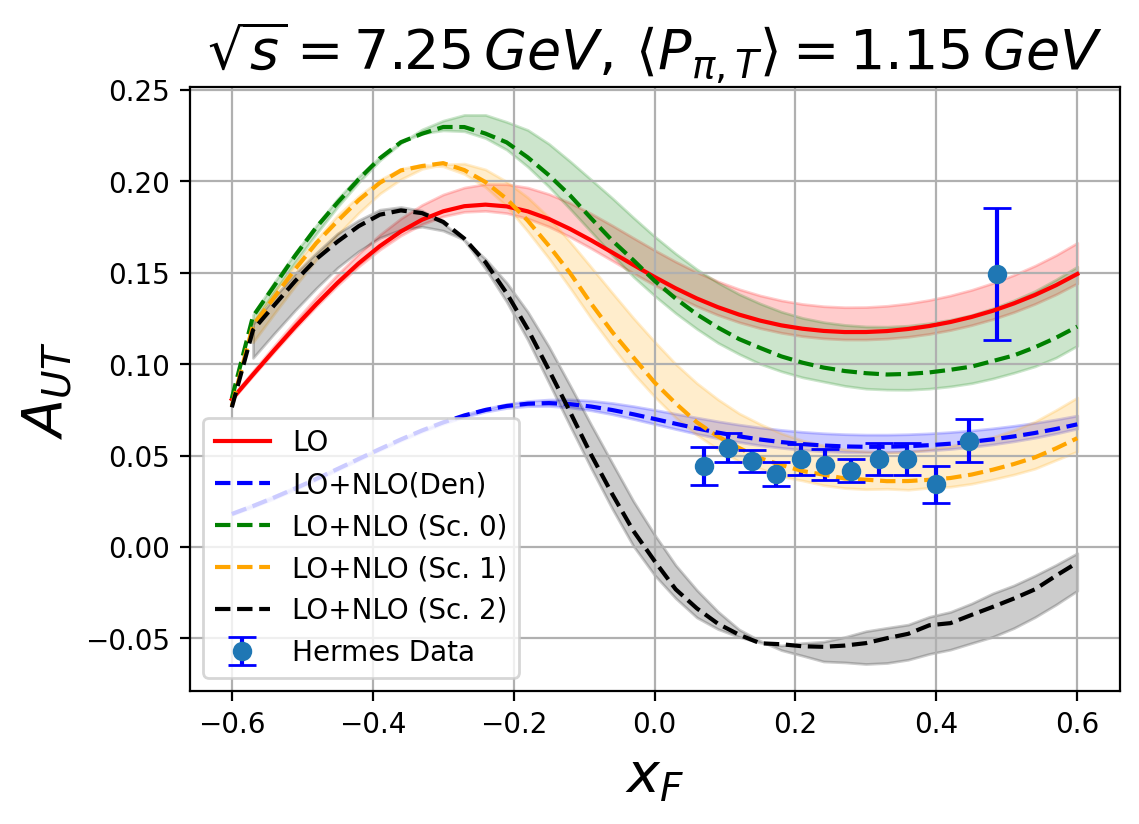,width=0.49\textwidth}
  \epsfig{figure=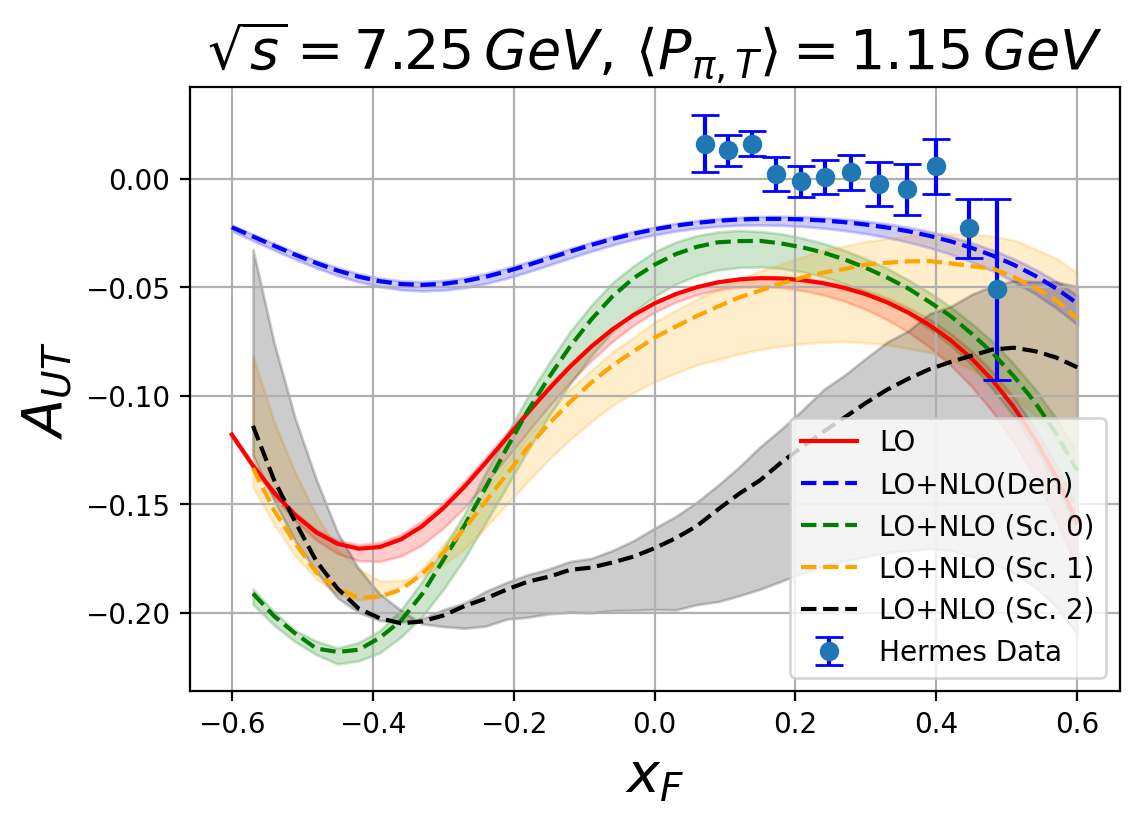,width=0.49\textwidth}
 \vspace*{0cm} 
  \caption{$A_{UT}$ at LO and NLO for our three scenarios of twist-3 proton matrix elements, compared to HERMES data \cite{Airapetian:2013bim}
  for $ep^\uparrow \to\pi^+ X$ (left) and $ep^\uparrow \to\pi^- X$ (right). \label{fig4}}
\end{figure*}

Figure~\ref{fig4} shows the asymmetries $A_{UT}$ for $\pi^+$ (left) and $\pi^-$ (right) production, as functions
of  the pion's Feynman variable $x_F$. We present the LO result as well as the NLO ones for our three scenarios. 
Our main finding is that the NLO asymmetry is highly sensitive to the set of twist-3 matrix elements chosen, and the preliminary scale uncertainty bands do not overlap over a wide range in $x_F$.
We found that this is primarily due to an intricate interplay among the various pole contributions. 
We note that -- arguably accidentally -- our scenario~1 actually describes the HERMES $\pi^+$ data rather well, while none of
the scenarios does a particularly good job for $\pi^-$. We stress how crucial our NLO corrections are in this context, and 
how complex the twist-3 framework is: by construction, the LO result -- which only probes the diagonal $x=x'$ -- 
is the same for each of the three model sets of twist-3 matrix elements. For further comparison, we also show
the asymmetry computed with NLO corrections only in the denominator (dashed lines). Thanks to sizable
positive NLO corrections for the unpolarized cross section it is strongly reduced relative to LO, as
was first observed in~\cite{Fitzgibbons:2024zsh}. Our results demonstrate that inclusion of NLO
corrections also in the numerator of $A_{UT}$ is crucial for phenomenology. 

\textbf{\textit{Conclusions.}} We have presented the NLO QCD corrections to the spin-dependent 
cross section for the single-inclusive process $\ell p^\uparrow\to hX$ with transversely polarized protons. 
We have demonstrated that collinear twist-3 factorization holds at the one-loop level for the 
contributions associated with twist-3 nucleon matrix elements. This is a highly non-trivial result,
given the complexity of the twist-3 framework for polarized {\it single-inclusive} cross sections, compared to that for the Drell-Yan process and semi-inclusive DIS. In fact, this is the
first time that it has been possible to obtain the NLO corrections for a single-inclusive single-spin asymmetry with its all-important derivative terms.
Our calculation thus serves as a proof of principle that such NLO twist-3 calculations 
are possible, including phenomenological evaluations. 
The NLO single-spin asymmetry $A_{UT}$ shows great sensitivity to the set of twist-3 matrix elements
adopted, which also highlights the impact that future precision EIC data might have. We hope that extended data coverage at the EIC will make it possible to better constrain the 
twist-3 correlation functions also for $x\neq x'$ so that one can go 
beyond just modeling this regime. This, along with full use of evolution of the functions, will remain a challenge for the future.
In any case, our work presented here opens the door to future full calculations of $pp^\uparrow\to hX$.


  

\twocolumngrid

\textbf{\textit{Acknowledgements.}} 
W.V. is grateful to Patriz Hinderer and Yuji Koike for their collaboration at initial stages of this work. 
We thank Jianwei Qiu for helpful discussions. This work has been supported by Deutsche 
Forschungsgemeinschaft (DFG) through the Research Unit FOR 2926 (Project No. 409651613).

\bibliography{refs}

\end{document}